\documentclass[leqno, 10pt]{article}
\usepackage{times}
\usepackage{amsmath}
\usepackage{amssymb}
\usepackage{setspace}

\numberwithin{equation}{section}

\newtheorem{theorem}{Theorem}[section]
\newtheorem{definition}[theorem]{Definition}
\newtheorem{lemma}[theorem]{Lemma}
\newtheorem{corollary}[theorem]{Corollary}

\begin{document}

\date{}

\title{Algebraic semantics for a modal logic close to S1}

\author{Steffen Lewitzka\thanks{Universidade Federal da Bahia -- UFBA,
Instituto de Matem\'atica,
Departamento de Ci\^encia da Computa\c c\~ao,
Campus de Ondina,
40170-110 Salvador -- BA,
Brazil,
e-mail: steffen@dcc.ufba.br}}

\maketitle
\begin{abstract}
The modal systems S1--S3 were introduced by C. I. Lewis as logics for strict implication. While there are Kripke semantics for S2 and S3, there is no known natural semantics for S1. We extend S1 by a Substitution Principle SP which generalizes a reference rule of S1. In system S1+SP, the relation of strict equivalence $\varphi\equiv\psi$ satisfies the identity axioms of R. Suszko's non-Fregean logic adapted to the language of modal logic (we call these axioms the axioms of propositional identity). This enables us to develop a framework of algebraic semantics which captures S1+SP as well as the Lewis systems S3--S5. So from the viewpoint of algebraic semantics, S1+SP turns out to be an interesting modal logic. We show that S1+SP is strictly contained between S1 and S3 and differs from S2. It is the weakest modal logic containing S1 such that strict equivalence is axiomatized by propositional identity.
\end{abstract}

Keywords: modal logic, strict equivalence, propositional identity, non-Fregean logic, algebraic semantics

\section{Introduction}

Discontented with the notion of material implication found in the \textit{Principia Mathematica}, C. I. Lewis introduced axiomatizations of \textit{strict} implication \cite{lew, lewlan}. These systems are known as the non-normal modal logics S1, S2 and S3. In \cite{lewlan} also appeared (deductively equivalent systems of) the modal logics S4 and S5 which, however, are not accepted by Lewis as systems for strict implication. Possible worlds semantics for modal logics was introduced years later by S. Kripke and has become the standard semantics. Normal Kripke semantics involves rather strong modal principles such as the Necessitation Rule. Some systems which do not validate the Necessitation Rule, such as S2 and S3, can be captured by Kripke semantics if one is ready to accept the concept of a \textit{non-normal world}. The semantical treatment of other logics in the vicinity of S1, however, seems to be more complicated. In fact, the known semantics for S1 (M. J. Cresswell \cite{cre3, cre4}, see also R. Girle \cite{gir1, gir2}) and related systems (see, e.g., B. F. Chellas and K. Segerberg \cite{cheseg}) are technically more complex and mostly unintuitive.

In this paper, we are inspired by ideas coming from R. Suszko's non-Fregean logic (\cite{blosus, sus}, see also \cite{mal}). The essential feature of a non-Fregean logic is an identity connective $\equiv$ such that formulas of the form $(\varphi\equiv\psi)\rightarrow (\varphi\leftrightarrow\psi)$ are theorems but the so-called Fregean Axiom $ (\varphi\leftrightarrow\psi)\rightarrow (\varphi\equiv\psi)$ is not a theorem. $\varphi\equiv\psi$ can be read as ``$\varphi$ and $\psi$ have the same meaning" or ``$\varphi$ and $\psi$ denote the same proposition".\footnote{Motivated by Wittgenstein's \textit{Tractatus}, Suszko \cite{susz} applies the term \textit{situation} instead of \textit{proposition}. By a \textit{proposition} we mean the denotation of a sentence (a formula) in a non-Fregean model under a given valuation. In modal logic, a proposition is usually defined as a set of possible worlds. Both views on a proposition are very similar (see, e.g., \cite{lewarx, pol}).} If one forces the Fregean Axiom to be valid, then the underlying non-Fregean logic specializes to classical logic where models contain only two propositions: the True and the False; that is, a proposition is reduced to its truth-value.\footnote{We call such models extensional. Among others, there are also intensional models where a proposition not only represents a truth-value but also the intension of a sentence (see, e.g., \cite{lewigpl}).} 

Suszko, in his non-Fregean approach to modality \cite{sus,blosus}, introduces necessity by the equational axiom scheme $\square\varphi\equiv(\varphi\equiv\top)$, where the tautological formula $\top$ denotes \textit{the necessary proposition (situation)}. He defines two theories $W_T$ and $W_H$ whose models are certain topological Boolean algebras, also called interior algebras in the literature. Applying a classical result due to McKinsey and Tarski, Suszko then is able to conclude that the theories $W_T$ and $W_H$ correspond to the modal logics S4 and S5, respectively. Similar axiomatic approaches, where the modal operator $\square$ is introduced by means of a given identity connective, were studied by Cresswell \cite{cre1, cre2}, Martens \cite{mar} and other authors; see also the historical note at the end of \cite{sus}.

In contrast to Suszko's approach, we shall work in this paper with the pure language of modal logic (i.e., the language of propositional logic augmented with a modal operator $\square$). Instead of defining modality by means of an identity connective, we go the other way arround and define an ``identity connective" by strict equivalence: $\varphi\equiv\psi := \square(\varphi\rightarrow\psi)\wedge\square(\psi\rightarrow\varphi)$. We require that the so defined connective satisfies what we call here \textit{the axioms of propositional identity}. These are the axioms that result from the adaptation of the identity axioms of basic non-Fregean logic ((e)--(h) of Definition 1.1. in \cite{blosus}) to the language of modal logic. It turns out that it is enough to consider Lewis modal system S1 extended by an inference rule SR which is slightly stronger than the S1-rule of \textit{Substitution of Proved Strict Equivalents} (see, e.g., \cite{hugcre}). Rule SR is equivalent with the \textit{Substitution Principle} SP of non-Fregean logic and ensures that $\equiv$ has the desired properties in the resulting modal system S1+SP.\footnote{Suszko's theories $W_T$ and $W_H$ \cite{sus} rely on stronger assumptions. The coincidence of Suszko's identity $\equiv$ with strict equivalence derives as a theorem (see (12.16) and the following Metatheorem VI on p. 21 in \cite{sus}): $(\varphi\equiv\psi)\equiv\square(\varphi\leftrightarrow\psi)$. In fact, the idea to identify propositional identity with strict equivalence is not only inherent in Suszko's work but is also present already in Cresswell's approach (see \cite{cre2, mar}).} This enables us to develop a non-Fregean semantics whose models are given by Boolean algebras with some additional structure. Our semantics seems to be similar to standard algebraic semantics for \textit{normal} modal logics which relies on Boolean algebras with operators, sometimes called \textit{modal algebras} in the literature (see, e.g., \cite{chazak}).\footnote{Note that the topological Boolean algebras used by Suszko \cite{sus} are special cases of such modal algebras.} A modal algebra is a Boolean algebra with an unary operation $f_\square$ (more than one operation can be considered) such that $f_\square(f_\top)=f_\top$ and $f_\square(f_\wedge(m,m'))=f_\wedge(f_\square(m),f_\square(m'))$, for all elements $m,m'$ (of course, $f_\top$ is the top element and $f_\wedge$ is the meet operation). These equations, however, are generally not satisfied in our models. Formulating certain semantic constraints, we obtain a framework of algebraic semantics for the sequence of modal systems S1+SP, S3, S4, S5. A nice feature of this framework is that normal and non-normal systems are captured in a conceptually uniform way. We obtain \textit{strong} completeness theorems. We also show that the inclusions S1$\subseteq$(S1+SP)$\subseteq$S3 are strict, and (S1+SP)$\neq$S2. S1+SP can be characterized as the weakest modal logic containing S1 such that the relation of strict equivalence is given by propositional identity.

\section{The modal logic S1+SP}

The set $Fm$ of formulas of modal propositional logic is defined in the usual way over a set $V=\{x_0,x_1,x_2,...\}$ of propositional variables, logical connectives $\neg, \rightarrow$ and the modal operator $\square$ for necessity. If $x$ is a variable and $\varphi,\psi$ are formulas, then we write $\varphi[x:=\psi]$ for the formula that results from substituting $\psi$ for all occurrences of $x$ in $\varphi$. We use the following abbreviations:
\begin{itemize}
\item $\top:=(x_0\rightarrow x_0)$, $\bot:=\neg\top$
\item $\varphi\wedge\psi := \neg(\varphi\rightarrow\neg\psi)$
\item $\varphi\leftrightarrow\psi :=(\varphi\rightarrow\psi)\wedge(\psi\rightarrow\varphi)$
\item $\varphi\equiv\psi := \square(\varphi\rightarrow\psi)\wedge\square(\psi\rightarrow\varphi)$
\end{itemize}

In particular, $\varphi\equiv\psi$ reads as strict equivalence of $\varphi$ and $\psi$. The following axiomatization of Lewis system S1 is due to E. J. Lemmon \cite{lem}. In contrast to Lewis original axiomatization found in \cite{lewlan}, Lemmon's axiomatization is formulated as an extension of the calculus of (non-modal) propositional logic. All formulas of the following form, and only those, are axioms:\\

\noindent (i) formulas which have the form of a propositional tautology\\
(ii) $\square\varphi\rightarrow\varphi$\\
(iii) $(\square(\varphi\rightarrow\psi)\wedge\square(\psi\rightarrow\chi))\rightarrow\square(\varphi\rightarrow\chi)$\\

The inference rules of S1 are
\begin{itemize}
\item Modus Ponens MP: ``From $\varphi$ and $\varphi\rightarrow\psi$ infer $\psi$."
\item Axiom Necessitation AN: ``If $\varphi$ is an axiom, then infer $\square\varphi$."
\item Substitution of Proved Strict Equivalents SPSE: ``If $\psi\equiv\psi'$ is a theorem, then any formula of the form $\varphi[x:=\psi]\equiv\varphi[x:=\psi']$ is a theorem." 
\end{itemize}

Lewis system S3 results from S1 by adding\\
\noindent (3) \centerline{$\square(\varphi\rightarrow\psi)\rightarrow\square(\square\varphi\rightarrow\square\psi)$} \\
\noindent as axiom scheme. Rule AN now applies to the axioms (i)--(iii) and also to (3). We write S3=S1+(3). We shall see that S2 cannot be captured by our semantic approach, so we do not give an axiomatization of that system here (see, e.g., \cite{hugcre} for a Lemmon-style axiomatization).

S1 proves to be sound with respect to the algebraic semantics presented below. However, it is not complete: rule SPSE is too weak. Therefore, we generalize SPSE to the following Substitution Rule SR:\\

``If $\chi\rightarrow(\psi\equiv\psi')$ is a theorem, then $\chi\rightarrow(\varphi[x:=\psi]\equiv\varphi[x:=\psi'])$ is a theorem."\\

\noindent Then formulas of the form

\noindent SP \centerline{$(\psi\equiv\psi')\rightarrow(\varphi[x:=\psi]\equiv\varphi[x:=\psi'])$}

\noindent are theorems: choose $\chi = (\psi\equiv\psi')$ in rule SR. On the other hand, modulo S1, rule SR derives from SP together with transitivity of implication. That is, SR and SP are equivalent modulo S1.

Scheme SP is what we call the Substitution Principle of non-Fregean logic (see also, e.g., \cite{lewigpl}, Lemma 3.3). Actually, it represents a general ontological law known as \textit{the Indiscernibility of Identicals}. In basic non-Fregean logic SCI, principle SP follows from the identity axioms (e)--(g) of Definition 1.1 in \cite{blosus}; see also the remark following Proposition 1.3 in \cite{blosus}. If we adapt Suszko's identity axioms to the modal language, then SP is established by the following axioms:\\

\noindent (iv) $(\varphi\equiv\psi)\rightarrow(\neg\varphi\equiv\neg\psi)$\\
(v)  $((\varphi\equiv\psi)\wedge(\varphi'\equiv\psi'))\rightarrow((\varphi\rightarrow\varphi')\equiv(\psi\rightarrow\psi'))$\\
(vi) $(\varphi\equiv\psi)\rightarrow(\square\varphi\equiv\square\psi)$\\
 
It is not hard to recognize that (iv)--(vi) are equivalent with scheme SP. That is, SP can be seen as an abbreviation of (iv)--(vi). In the following, we consider the modal system S1+SP which results from S1 by adding the axiom schemes (iv)--(vi) (and ignoring rule SPSE which is weaker than SP). As in S1, the inference rules are again MP and AN. However, the application of rule AN remains restricted to the axioms (i)--(iii) of the original system S1. A derivation of a formula $\varphi$ from a set of formulas $\Phi$ in system S1+SP is a finite sequence of formulas $\varphi_0,...,\varphi_n=\varphi$ such that for each $i=0,...,n$, formula $\varphi_i$ is either an element of $\Phi$, or it is an axiom, or it is the conclusion of rule AN applied to an axiom of type (i)--(iii), or it is the conclusion of rule MP with premises that appear as $\varphi_j$ and $\varphi_k=(\varphi_j\rightarrow\varphi_i)$ in the sequence with $j,k<j$. We write $\Phi\vdash\varphi$ if there exists a derivation of $\varphi$ from $\Phi$ in deductive system S1+SP. \\

If we let rule AN be applicable to all axioms (i)--(vi), then we get a stronger system which we denote by S1+$\square$SP. Both S1+SP and S1+$\square$SP are sublogics of S3 as we shall see below. 

If we consider the Lemmon-style axiomatization of system S2 such as given in \cite{hugcre}, then S3=S2+(3), where AN applies here to all axioms of S2 and also to (3). S2 contains a rule called \text{Becker's Rule}: ``If $\square(\varphi\rightarrow\psi)$ is a theorem, then $\square(\square\varphi\rightarrow\square\psi)$ is a theorem." Now observe that scheme (3) generalizes Becker's rule in exactly the same way as SP generalizes the S1-rule SPSE. That is, S1+$\square$SP corresponds to S1 in the way as S3 corresponds to S2. The logic S1+$\square$SP, however, is less important for our purposes. In the present paper, we are interested in a \textit{minimal} modal logic which can be integrated into the hierarchy of Lewis systems S1--S5 such that strict equivalence satisfies SP, i.e., the axioms (iv)--(vi). It turns out that S1+SP is strong enough for our purposes. So we will focus on that system.\\

The following Deduction Theorem can be shown by induction on the length of a derivation, similarly as in \cite{lewarx}, Lemma 2.3.

\begin{lemma}(Deduction Theorem)\label{90}
If $\Phi\cup\{\varphi\}\vdash\psi$, then $\Phi\vdash\varphi\rightarrow\psi$.
\end{lemma}

\begin{lemma}\label{92}
$\square\varphi\leftrightarrow(\varphi\equiv\top)$ is a theorem, for any formula $\varphi$.
\end{lemma}

\paragraph*{Proof.}
By rule AN we derive the following: $\square((\varphi\rightarrow\top)\rightarrow\top)$, $\square(\top\rightarrow (\varphi\rightarrow\top))$, $\square((\top\rightarrow\varphi)\rightarrow\varphi)$ and $\square(\varphi\rightarrow(\top\rightarrow\varphi))$. This yields\\ 
(a) $\vdash(\varphi\rightarrow\top)\equiv\top$ and \\
(b) $\vdash(\top\rightarrow\varphi)\equiv\varphi$.\\
Let $\chi_1$ be the formula $\square\varphi\leftrightarrow(\square x\wedge \square(\top\rightarrow\varphi))$, where $x$ is a fresh variable. By SP,\\
$\vdash((\varphi\rightarrow\top)\equiv\top) \rightarrow\chi_1[x:=(\varphi\rightarrow\top)]\equiv\chi_1[x:=\top])$. \\
Then (a) and MP yield \\
(c) $\vdash (\square\varphi\leftrightarrow(\square(\varphi\rightarrow\top)\wedge\square (\top\rightarrow\varphi)))\equiv (\square\varphi\leftrightarrow(\square\top\wedge\square(\top\rightarrow\varphi)))$. \\
Now let $\chi_2=\square\varphi\leftrightarrow(\square \top\wedge\square y)$, where $y$ is a fresh variable. By SP, \\
$\vdash((\top\rightarrow\varphi)\equiv\varphi)\rightarrow(\chi_2[y:=(\top\rightarrow\varphi)]\equiv\chi_2[y:=\varphi])$. \\
Then (b) and MP yield\\
(d) $\vdash (\square\varphi\leftrightarrow(\square\top\wedge\square(\top\rightarrow\varphi)))\equiv (\square\varphi\leftrightarrow(\square\top\wedge\square \varphi))$.\\
By axiom (iii), the identity connective is transitive. Thus, (c) and (d) imply\\
$\vdash(\square\varphi\leftrightarrow(\square(\varphi\rightarrow\top)\wedge\square (\top\rightarrow\varphi)))\equiv (\square\varphi\leftrightarrow(\square\top\wedge\square \varphi))$.\\
That is,\\
$\vdash (\square\varphi\leftrightarrow(\varphi\equiv\top))\equiv (\square\varphi\leftrightarrow(\square\top\wedge\square \varphi))$.\\
The formula on the right hand side of the last equation is obviously a theorem. Then the formula on the left hand side is a theorem, too. Q.E.D.\\

We shall refer to the scheme $\square\varphi\leftrightarrow(\varphi\equiv\top)$ as principle N. It says that a proposition denoted by $\varphi$ is necessary iff $\varphi$ and $\top$ denote the same proposition. In other words, ``There is exactly one necessary proposition".

\begin{lemma}\label{95}
Modal principle K is a theorem: $\vdash\square(\varphi\rightarrow\psi)\rightarrow(\square\varphi\rightarrow\square\psi)$.
\end{lemma}

\paragraph*{Proof.}
We have\\
$\{\square(\varphi\rightarrow\psi),\square\varphi\}\vdash\square(\varphi\rightarrow\psi)$\\ 
$\{\square(\varphi\rightarrow\psi),\square\varphi\}\vdash\square\varphi$\\
$\{\square(\varphi\rightarrow\psi),\square\varphi\}\vdash\varphi\equiv\top$, by premise $\square\varphi$, principle N and MP\\
$\{\square(\varphi\rightarrow\psi),\square\varphi\}\vdash(\varphi\equiv\top)\rightarrow (\square(x\rightarrow\psi)[x:=\varphi]\equiv\square(x\rightarrow\psi)[x:=\top])$, by SP, where $x$ is a fresh variable\\
$\{\square(\varphi\rightarrow\psi),\square\varphi\}\vdash(\varphi\equiv\top)\rightarrow (\square(\varphi\rightarrow\psi)\equiv\square(\top\rightarrow\psi))$\\
$\{\square(\varphi\rightarrow\psi),\square\varphi\}\vdash \square(\varphi\rightarrow\psi)\equiv\square(\top\rightarrow\psi)$, by MP\\
$\{\square(\varphi\rightarrow\psi),\square\varphi\}\vdash \square(\top\rightarrow\psi)$, by axiom scheme (ii) and MP\\
$\{\square(\varphi\rightarrow\psi),\square\varphi\}\vdash \square(\psi\rightarrow\top)$, by AN\\
$\{\square(\varphi\rightarrow\psi),\square\varphi\}\vdash \psi\equiv\top$\\
$\{\square(\varphi\rightarrow\psi),\square\varphi\}\vdash (\psi\equiv\top)\rightarrow\square\psi$, by principle N\\
$\{\square(\varphi\rightarrow\psi),\square\varphi\}\vdash \square\psi$, by MP\\
$\vdash\square(\varphi\rightarrow\psi)\rightarrow(\square\varphi\rightarrow\square\psi)$, by applying two times the Deduction Theorem. Q.E.D.

\section{Semantics}

A Boolean algebra is usually given by a non-empty universe together with operations for join, meet, complement and two distinguished elements (the smallest and the greatest element w.r.t. the induced lattice ordering) such that certain equations are satisfied. In the following definition we make use of the fact that, by interdefinability of operations, a Boolean algebra can be equivalently given by operations for complement and implication.

\begin{definition}\label{100}
A model $\mathcal{M}=(M, TRUE, f_\neg, f_\rightarrow, f_\square)$ is given by an universe $M$ of abstract entities, called propositions, a set $TRUE\subseteq M$ of true propositions and operations $f_\neg$, $f_\rightarrow$ for complement and implication which form a Boolean algebra on $M$, and an unary operation $f_\square$ such that for all $m,m',m''\in M$ the following conditions (i)--(vi) are satisfied. By $\le$ we denote the induced lattice ordering, $f_\top, f_\bot$ are the greatest, the smallest element of the lattice, respectively, and $f_\wedge$ is the meet operation of the Boolean algebra. 
\begin{enumerate}
\item $f_\bot\notin TRUE$, $f_\top\in TRUE$
\item $f_\neg(m)\in TRUE\Leftrightarrow m\notin TRUE$
\item $f_\rightarrow(m,m')\in TRUE\Leftrightarrow m\notin TRUE$ or $m'\in TRUE$
\item $f_\square(m)\in TRUE\Leftrightarrow m=f_\top$
\item $f_\square(m)\le m$
\item $f_\wedge(f_\square(f_\rightarrow(m,m')),f_\square(f_\rightarrow(m',m'')))\le f_\square(f_\rightarrow(m,m''))$
\end{enumerate}
\end{definition}

As every Boolean algebra, a model $\mathcal{M}$ satisfies $m\le m'\Leftrightarrow f_\rightarrow(m,m')=f_\top$. We will make tacitly use of this fact in some of the proofs below. The conditions (i)--(iii) in the definition ensure that $TRUE$ is an ultrafilter of the Boolean lattice. The conditions (v) and (vi) are semantic counterparts of applications of rule AN to the axioms (ii) and (iii), respectively. Condition (iv) specifies the properties of the modal operator. Later, we will strengthen that condition in order to obtain semantics for stronger modal logics, namely for the Lewis systems S3--S5.

\begin{definition}\label{120}
Let $\mathcal{M}$ be a model. An assignment of propositions to formulas is a function $\gamma:V\rightarrow M$ which extends in the canonical way to a function $\gamma:Fm\rightarrow M$ (we refer to the extension again by $\gamma$).\footnote{An assignment is sometimes called a valuation.} That is, $\gamma(\neg\varphi)=f_\neg(\gamma(\varphi))$, $\gamma(\varphi\rightarrow\psi)=f_\rightarrow(\gamma(\varphi),\gamma(\psi))$ and $\gamma(\square\varphi)=f_\square(\gamma(\varphi))$. If $\mathcal{M}$ is a model and $\gamma$ is a corresponding assignment, then we call the tuple $(\mathcal{M},\gamma)$ an interpretation. The satisfaction relation between interpretations and formulas is defined by $(\mathcal{M},\gamma)\vDash\varphi :\Leftrightarrow\gamma(\varphi)\in TRUE$. This definition extends in the usual way to sets of formulas. For a set of formulas $\Phi$ we define $Mod(\Phi)=\{(\mathcal{M},\gamma)\mid (\mathcal{M},\gamma)\vDash\Phi\}$. The relation of logical consequence is defined by: $\Phi\Vdash\varphi :\Leftrightarrow Mod(\Phi)\subseteq Mod(\{\varphi\})$.
\end{definition}

Now it is clear how to interpret the lattice ordering $\le$ of a given model $\mathcal{M}$. If $m,m'\in M$, then $m\le m'$ means that proposition $m$ strictly implies proposition $m'$.\footnote{We have $m\le m'$ iff $f_\rightarrow(m,m')=f_\top$ iff $f_\square(f_\rightarrow(m,m'))\in TRUE$. Choose variables $x,y$ and an assignment $\gamma$ such that $\gamma(x)=m$ and $\gamma(y)=m'$. Then $(\mathcal{M,\gamma})\vDash \square(x\rightarrow y)\Leftrightarrow m\le m'$.}

The following result says that the defined connective $\equiv$ has the intended semantics, i.e., it is an identity connective:

\begin{theorem}\label{124}
If $(\mathcal{M},\gamma)$ is an interpretation and $\varphi,\psi\in Fm$, then 
\begin{equation*}
(\mathcal{M},\gamma)\vDash\varphi\equiv\psi\Leftrightarrow\gamma(\varphi)=\gamma(\psi).
\end{equation*}
\end{theorem} 

\paragraph*{Proof.}
Suppose $(\mathcal{M},\gamma)\vDash\varphi\equiv\psi$. That is, $(\mathcal{M},\gamma)\vDash\square(\varphi\rightarrow\psi)\wedge\square(\psi\rightarrow\varphi)$. Let $\gamma(\varphi)=m$ and $\gamma(\psi)=m'$. Then follows that $f_\square(f_\rightarrow(m,m'))\in TRUE$ and $f_\square(f_\rightarrow(m',m))\in TRUE$. Thus, $m\le m'$ and $m'\le m$, and finally $m=m'$. On the other hand, if $\gamma(\varphi)=m=\gamma(\psi)$, then $f_\square(f_\rightarrow(m,m))\in TRUE$, since $m\le m$. It follows that $(\mathcal{M},\gamma)\vDash\varphi\equiv\psi$. Q.E.D. 

\begin{corollary}\label{126}
Principle SP is valid: $\Vdash(\psi\equiv\psi')\rightarrow(\varphi[x:=\psi]\equiv\varphi[x:=\psi'])$.
\end{corollary}

\paragraph*{Proof.}
Suppose $(\mathcal{M},\gamma)\vDash\psi\equiv\psi'$. By Theorem \ref{124}, $\gamma(\psi)=\gamma(\psi')$. By induction on $\varphi$ it follows that $\gamma(\varphi[x:=\psi])=\gamma(\varphi[x:=\psi'])$. Theorem \ref{124} implies $(\mathcal{M},\gamma)\vDash\varphi[x:=\psi]\equiv\varphi[x:=\psi']$. Q.E.D.\\

This result also provides the following semantical interpretation of SP. If $\psi,\psi'$ and $\varphi$ are formulas such that $\psi$ is a subformula of $\varphi$, and $\psi$ has the same denotation as $\psi'$, then $\varphi$ has the same denotation as $\varphi'$, where $\varphi'$ is the result of replacing $\psi$ in $\varphi$ by $\psi'$.\\

Our next goal is to prove that S1+SP is sound. Let $\mathcal{M}$ be a model with universe $M$. With a given assignment $ß\gamma:V\rightarrow M$ we may associate a truth--value assignment $\beta_\gamma:V\rightarrow\{0,1\}$ defined by $\beta_\gamma(x)=1 :\Leftrightarrow\gamma(x)\in TRUE$. It follows that for all non-modal propositional formulas $\varphi$, $\gamma(\varphi)\in TRUE\Leftrightarrow\beta_\gamma(\varphi)=1$. Thus, $(\mathcal{M},\gamma)\vDash\varphi$, for all propositional tautologies $\varphi$. Now suppose $\varphi$ is a modal formula having the form of a propositional tautology, i.e. $\varphi$ is the result of replacements of variables by modal formulas within a propositional tautology. By soundness of SP, Corollary \ref{126}, $\varphi$ has the same denotation as the original tautology. This holds in every model, thus $\varphi$ is valid. Since $f_\top\in TRUE$, axiom (ii) is valid. Axiom (iii) says that the lattice ordering of any model is transitive. We have shown that the S1-axioms (i)--(iii) are valid. The validity of (iv)--(vi) is guaranteed by Corollary \ref{126}. Obviously, rule MP is sound. It is well-known that in any Boolean algebra, any assignment maps all propositional tautologies to the top element of the lattice. Thus, if $\varphi$ is an axiom of the form (i), then $\square\varphi$ is valid. The conditions (v) and (vi) of Definition \ref{100} ensure that the same holds for axioms of the form (ii) and (iii), respectively. Thus, rule AN is sound (recall that we have restricted the application of AN to axioms of the form (i)--(iii)). The soundness of the deductive system now follows by induction on the length of a derivation.

\begin{theorem}[Soundness]\label{140}
For any set $\Phi\cup\{\varphi\}\subseteq Fm$: $\Phi\vdash\varphi\Rightarrow\Phi\Vdash\varphi$.
\end{theorem}

\section{The Completeness Theorem}

In order to prove strong completeness of S1+SP we follow the usual strategy. We call a set of formulas consistent if there is a formula which is not derivable from that set. A set which is not consistent is called inconsistent. A maximal consistent set of formulas is a consistent set such that every proper extension is inconsistent. From standard arguments it follows that each consistent set extends to a maximal consistent set. It remains to show that a maximal consistent set has a model. We will tacitly make use of the following well-known properties of a maximal consistent set $\Phi$. These properties follow from propositional logic:
\begin{itemize}
\item $\varphi\in\Phi\Leftrightarrow\Phi\vdash\varphi$
\item $\neg\varphi\in\Phi\Leftrightarrow\varphi\notin\Phi$
\item $\varphi\rightarrow\psi\in\Phi\Leftrightarrow\varphi\notin\Phi$ or $\psi\in\Phi$.
\end{itemize}

\begin{definition}\label{300}
For a maximal consistent set $\Phi$ we define a relation $\approx_\Phi$ on $Fm$ by $\varphi\approx_\Phi\psi :\Leftrightarrow\Phi\vdash\varphi\equiv\psi$.
\end{definition}

\begin{lemma}\label{320}
Let $\Phi$ be a maximal consistent set. The relation $\approx_\Phi$ is an equivalence relation on $Fm$ with the following properties:
\begin{itemize}
\item If $\varphi_1\approx_\Phi\psi_1$ and $\varphi_2\approx_\Phi\psi_2$, then $\neg\varphi_1\approx_\Phi\neg\psi_1$, $\square\varphi_1\approx_\Phi\square\psi_1$ and $\varphi_1\rightarrow\varphi_2\approx_\Phi\psi_1\rightarrow\psi_2$.
\item If $\varphi\approx_\Phi\psi$, then $\varphi\in\Phi\Leftrightarrow\psi\in\Phi$.
\item If $\varphi\approx_\Phi\psi$, then $\square\varphi\in\Phi\Leftrightarrow\square\psi\in\Phi$.
\end{itemize}
\end{lemma}

\paragraph*{Proof.}
Symmetry of $\approx_\Phi$ follows from propositional logic. Since $\varphi\rightarrow\varphi$ is an axiom, we get $\square(\varphi\rightarrow\varphi)\in\Phi$ by rule AN. Thus, $\approx_\Phi$ is reflexive. Transitivity follows from the scheme of axioms (iii). Now suppose $\varphi_1\approx_\Phi\psi_1$ and $\varphi_2\approx_\Phi\psi_2$. Let $x\neq y$ be variables such that $x$ does not occur in $\psi_2$ and $y$ does not occur in $\varphi_1$. Then by SP and MP: $(\varphi_1\rightarrow\varphi_2)=(\varphi_1\rightarrow y)[y:=\varphi_2]\approx_\Phi(\varphi_1\rightarrow y)[y:=\psi_2]=(\varphi_1\rightarrow\psi_2)=(x\rightarrow\psi_2)[x:=\varphi_1]\approx_\Phi(x\rightarrow\psi_2)[x:=\psi_1]=(\psi_1\rightarrow\psi_2)$. The remaining cases of the first item of the Lemma follow similarly. The second item of the Lemma follows from the scheme of axioms (ii). Finally, the third item follows from the previous assertions of the Lemma. Q.E.D.

\begin{lemma}\label{340}
Any maximal consistent has a model.
\end{lemma}

\paragraph*{Proof.}
Let $\Phi$ be a maximal consistent set of formulas. By $\overline{\varphi}$ we denote the equivalence class of $\varphi\in Fm$ modulo $\approx_\Phi$. We are going to construct a model with the following ingredients:
\begin{itemize}
\item $M:=\{\overline{\varphi}\mid\varphi\in Fm\}$ 
\item $TRUE:=\{\overline{\varphi}\mid\varphi\in \Phi\}$
\item functions $f_\neg$, $f_\rightarrow$,$f_\square$ defined by $f_\neg(\overline{\varphi}):=\overline{\neg\varphi}$, $f_\rightarrow(\overline{\varphi},\overline{\psi}):=\overline{\varphi\rightarrow\psi}$, $f_\square(\overline{\varphi}):=\overline{\square\varphi}$, respectively
\end{itemize}

By Lemma \ref{320}, all these ingredients are well-defined. We may define operations for join $f_\vee$, meet $f_\wedge$, top element $f_\top$ and bottom element $f_\bot$ by means of the given operations $f_\neg$ and $f_\rightarrow$. In order to prove that the defined operations form a Boolean algebra on $M$ it suffices to show that the elements of $M$ satisfy the equations which axiomatize the class of Boolean algebras. As an example, we choose the commutativity of the meet operation: $f_\wedge(\overline{\varphi},\overline{\psi})=f_\wedge(\overline{\psi},\overline{\varphi})$, for any $\overline{\varphi},\overline{\psi}\in M$. Since $(\varphi\wedge \psi)\leftrightarrow (\psi\wedge \varphi)$ is a tautology of propositional logic, rule AN implies $\Phi\vdash\square((\varphi\wedge \psi)\rightarrow (\psi\wedge \varphi))\wedge\square ((\psi\wedge \varphi)\rightarrow (\varphi\wedge \psi))$. That is, $\Phi\vdash (\varphi\wedge \psi)\equiv(\psi\wedge \varphi)$. Thus, $f_\wedge(\overline{\varphi},\overline{\psi})=\overline{\varphi\wedge\psi}=\overline{\psi\wedge\varphi}=f_\wedge(\overline{\psi},\overline{\varphi})$. Similarly, the remaining equational axioms of Boolean algebras are verified via propositional tautologies and rule AN. It is also clear that (i)--(iii) of Definition \ref{100} of a model are satisfied. Condition (iv) follows from Lemma \ref{92}. Finally, the conditions (v) and (vi) of a model are guaranteed by rule AN. Let $\mathcal{M}$ be the constructed model. We consider the canonical assignment $\gamma:V\rightarrow M$ defined by $x\mapsto\overline{x}$. By induction on $\varphi$ one shows that $\gamma(\varphi)=\overline{\varphi}$, for any formula $\varphi$. Then $\varphi\in\Phi\Leftrightarrow\overline{\varphi}\in TRUE\Leftrightarrow\gamma(\varphi)\in TRUE\Leftrightarrow (\mathcal{M},\gamma)\vDash\varphi$. Thus, $(\mathcal{M},\gamma)\vDash\Phi$. Q.E.D.

\begin{corollary}[Completeness Theorem]\label{360}
$\Phi\Vdash\varphi\Rightarrow\Phi\vdash\varphi$.
\end{corollary}

\paragraph*{Proof.}
$\Phi\nvdash\varphi$ implies the consistency of $\Phi\cup\{\neg\varphi\}$. We extend this set to a maximal consistent set which, by Lemma \ref{340}, has a model. Consequently, $\Phi\nVdash\varphi$. Q.E.D.

\section{Conclusions}

We saw that replacing rule SPSE with the stronger principle SP in modal logic S1 results in a non-normal modal logic which has a natural algebraic semantics. In this section, we discuss the relationships between S1+SP and the Lewis systems S1--S3. We also show that our semantics extends straightforwardly to semantics for the Lewis systems S3--S5, given by certain \textit{modal algebras}. As a negative result, we shall see that neither S1 nor S2 can be captured by our semantics.

The Substitution Principle SP, which is equivalent with the axioms (iv)--(vi), ensures that the identity connective, defined as strict equivalence, satisfies the following identity axioms: \\ 

\noindent (a) $\varphi\equiv\varphi$\\
(b) $(\varphi\equiv\psi)\rightarrow(\varphi\rightarrow\psi)$\\
(c) $(\varphi\equiv\psi)\rightarrow (\neg\varphi\equiv\neg\psi)$\\
(d) $((\varphi\equiv\psi)\wedge (\varphi'\equiv\psi'))\rightarrow ((\varphi\rightarrow\varphi')\equiv (\psi\rightarrow\psi'))$\\
(e) $(\varphi\equiv\psi)\rightarrow (\square\varphi\equiv\square\psi)$\\

Obviously, (a)--(e) are an adaptation of Suszko's identity axioms of non-Fregean logic \cite{blosus} to the language of modal logic. We call (a)--(e) the axioms of propositional identity. Note that (c)--(e) are precisely the axioms (iv)--(vi) of S1+SP. (a) and (b) are theorems of all Lewis modal systems: in order to obtain (a) apply AN, (b) follows from axiom (ii). Having in mind Suszko's non-Fregean approach to modal logic \cite{sus}, it is not surprising that (c)--(e) are theorems of S4 and S5. This leads to the following\\

\textbf{Question}: Are (c)--(e) theorems of all Lewis modal systems?\\

This is anwered by the next result which is taken from \cite{lew}, Theorem 5.3. For the convenience of the reader, we reproduce here the proof.

\begin{theorem}[\cite{lew}]\label{400}
In modal logic S3, the relation $\equiv$ of strict equivalence satisfies the axioms of propositional identity. That is, (a)--(e) above are derivable in S3. Moreover, S3 is the weakest Lewis modal logic with this property.
\end{theorem}

\paragraph*{Proof.}
As before, the connective $\equiv$ is defined as strict equivalence, i.e., $\varphi\equiv\psi := \square(\varphi\rightarrow\psi)\wedge\square(\psi\rightarrow\varphi)$. We have to show that under this assumption the axioms of propositional identity are derivable in S3. As already mentioned above, (a) and (b) are theorems of any Lewis system. So we concentrate on (c)--(e). First, we observe that $(\square\varphi\wedge\square\psi)\leftrightarrow\square(\varphi\wedge\psi)$ is a theorem of S3 and strict equivalence between $\varphi$ and $\psi$ can be expressed by $\square(\varphi\leftrightarrow\psi)$.\footnote{The same holds true in S1+SP. This follows from applications of modal principle K.} By rule AN we get $\square((\varphi\leftrightarrow\psi)\rightarrow(\neg\varphi\leftrightarrow\neg\psi))$. By MP and axiom K we obtain (c). Similarly, we may derive (d). Now we consider $\square(\varphi\rightarrow\psi)\rightarrow\square(\square\varphi\rightarrow\square\psi)$ and $\square(\psi\rightarrow\varphi)\rightarrow\square(\square\psi\rightarrow\square\varphi)$, which are axioms of S3. We derive $(\square(\varphi\rightarrow\psi)\wedge\square(\psi\rightarrow\varphi)) \rightarrow (\square(\square\varphi\rightarrow\square\psi)\wedge\square(\square\psi\rightarrow\square\varphi))$ and obtain (e). We have shown the first assertion of the Theorem. Note that axiom (e) is the scheme: $(\square(\varphi\rightarrow\psi)\wedge\square(\psi\rightarrow\varphi))\rightarrow (\square(\square\varphi\rightarrow\square\psi)\wedge\square(\square\psi\rightarrow\square\varphi))$. This formula, however, is not a theorem of S2. This can be shown by constructing a Kripke model of S2, i.e., a Kripke model with at least one normal world and reflexive accessibility relation, where that formula is not satisfied (hint: consider a model with exactly three normal worlds and an accessibility relation which is not transitive). Hence, in S1 and S2, strict equivalence cannot be the relation of propositional identity axiomatized by (a)--(e). Q.E.D.\\

By Theorem \ref{400}, principle SP is valid in S3 but not in S2. We conclude:

\begin{corollary}\label{410}
S1+SP is a sublogic of S3 and differs from S2.
\end{corollary}

\begin{lemma}\label{420}
Formulas of the following form are theorems of S3:
\begin{itemize}
\item $\square((\varphi\equiv\psi)\rightarrow (\neg\varphi\equiv\neg\psi))$
\item $\square(((\varphi\equiv\psi)\wedge (\varphi'\equiv\psi'))\rightarrow ((\varphi\rightarrow\varphi')\equiv (\psi\rightarrow\psi')))$
\item $\square((\varphi\equiv\psi)\rightarrow (\square\varphi\equiv\square\psi))$.
\end{itemize}
\end{lemma}

\paragraph*{Proof.}
We will use

\noindent (3) \centerline{$\square(\varphi\rightarrow\psi)\rightarrow\square(\square\varphi\rightarrow\square\psi)$}

\noindent which is an axiom scheme of S3. By AN, $\square((\varphi\leftrightarrow\psi)\rightarrow(\neg\varphi\leftrightarrow\neg\psi))$. By (3), $\square((\varphi\leftrightarrow\psi)\rightarrow(\neg\varphi\leftrightarrow\neg\psi))\rightarrow\square(\square(\varphi\leftrightarrow\psi)\rightarrow\square(\neg\varphi\leftrightarrow\neg\psi))$. Then by MP, $\square(\square(\varphi\leftrightarrow\psi)\rightarrow\square(\neg\varphi\leftrightarrow\neg\psi))$. This is the first theorem given in the lemma. By AN, $\square(((\varphi\leftrightarrow\psi)\wedge (\varphi'\leftrightarrow\psi'))\rightarrow ((\varphi\rightarrow\varphi')\leftrightarrow (\psi\rightarrow\psi')))$. By (3), $\square(((\varphi\leftrightarrow\psi)\wedge (\varphi'\leftrightarrow\psi'))\rightarrow ((\varphi\rightarrow\varphi')\leftrightarrow (\psi\rightarrow\psi')))\rightarrow \square(\square ((\varphi\leftrightarrow\psi)\wedge (\varphi'\leftrightarrow\psi'))\rightarrow \square((\varphi\rightarrow\varphi')\leftrightarrow (\psi\rightarrow\psi')))$. By MP, $\square(\square ((\varphi\leftrightarrow\psi)\wedge (\varphi'\leftrightarrow\psi'))\rightarrow \square((\varphi\rightarrow\varphi')\leftrightarrow (\psi\rightarrow\psi')))$. This is equivalent with $\square(\square(\varphi\leftrightarrow\psi)\wedge \square(\varphi'\leftrightarrow\psi'))\rightarrow \square((\varphi\rightarrow\varphi')\leftrightarrow (\psi\rightarrow\psi')))$. This is the second theorem given in the lemma. We put $\chi_1:=\square(\varphi\rightarrow\psi)$, $\chi_2:=\square(\psi\rightarrow\varphi)$, $\delta_1:=\square(\square\varphi\rightarrow\square\psi)$, $\delta_2:=\square(\square\psi\rightarrow\square\varphi)$. Rule AN applied to (3) yields $\square(\chi_1\rightarrow\delta_1)$ and $\square(\chi_2\rightarrow\delta_2)$. Thus, $\square(\chi_1\rightarrow\delta_1)\wedge\square(\chi_2\rightarrow\delta_2)$ is derivable in S3, and therefore also $\square((\chi_1\rightarrow\delta_1)\wedge(\chi_2\rightarrow\delta_2))$. By AN we also derive $\square(((\chi_1\rightarrow\delta_1)\wedge (\chi_2\rightarrow\delta_2))\rightarrow ((\chi_1\wedge\chi_2)\rightarrow (\delta_1\wedge\delta_2)))$. Now axiom K and rule MP yield $\square((\chi_1\wedge\chi_2)\rightarrow (\delta_1\wedge\delta_2))$. This is the third theorem in the lemma. Q.E.D.

\begin{corollary}
S1+$\square$SP is a sublogic of S3 (and differs from S2).
\end{corollary}

In the following we show that S1+SP differs from S3. Hence, it is a proper sublogic of S3. Recall that

\noindent (3) \centerline{$\square(\varphi\rightarrow\psi)\rightarrow\square(\square\varphi\rightarrow\square\psi)$}

\noindent is a theorem of S3. Interpreted in our semantics, (3) says that $f_\square$ is a monotonic operation: $m\le m'\Rightarrow f_\square(m)\le f_\square(m')$. We construct a model where $f_\square$ is not monotonic. Consider the powerset algebra of the set $\{1,2\}$ with the ultra-filter $TRUE=\{\{1,2\}, \{1\}\}$ and the operation $f_\square$ defined in the following way: $f_\square(\{1,2\})=\{1\}$, $f_\square(\{2\})=\{2\}$ and $f_\square(\{1\})=f_\square(\varnothing)=\varnothing$. Of course, the lattice ordering is given by set-theoretical inclusion. We have to show that this Boolean algebra is a model in the sense of Definition \ref{100}. One easily checks that the conditions (i)--(v) are satisfied. In order to verify condition (vi) it suffices to consider the following cases:\\
(A) $f_\square(f_\rightarrow(m,m'))=\{1\}$. This implies $f_\rightarrow(m,m')=\{1,2\}=f_\top$. That is, $m\le m'$.\\
(B) $f_\square(f_\rightarrow(m,m'))=\{2\}$. This implies $f_\rightarrow(m,m')=\{2\}$. Thus, ($m=\{1\}$ and $m'=\varnothing$) or ($m=\{1,2\}$ and $m'=\{2\}$).\\
(A)' $f_\square(f_\rightarrow(m',m''))=\{1\}$. This implies $m'\le m''$.\\
(B)' $f_\square(f_\rightarrow(m',m''))=\{2\}$. This implies ($m'=\{1\}$ and $m''=\varnothing$) or ($m'=\{1,2\}$ and $m''=\{2\}$).\\
Then it is enough to check the following two combinations: (A)+(A)', (B)+(B)'. The combination (B)+(B)' is inconsistent. The remaining combination (A)+(A)' implies $m\le m''$. This is equivalent with $f_\rightarrow(m,m'')=f_\top=\{1,2\}$. Consequently, $f_\square(f_\rightarrow(m,m''))=f_\square(\{1,2\})=\{1\}$ and condition (vi) holds true. Now we observe that $\{2\}\subseteq\{1,2\}$ and $f_\square(\{2\})\nsubseteq f_\square(\{1,2\}$. Thus, the operation $f_\square$ is not monotonic and scheme (3) is not valid in our semantics. Then (3) cannot be a theorem of S1+SP.

\begin{corollary}
S1+SP is a proper sublogic of S3.
\end{corollary}

\begin{corollary}
S1+SP is the smallest modal logic S such that S contains S1 and, modulo S, strict equivalence $\equiv$ satisfies the axioms of propositional identity (a)--(e) above. In this sense, modal logic S1+SP can be seen as the smallest non-Fregean theory that contains modal logic S1.
\end{corollary} 

From Lemma \ref{420} it follows that we may write S3=S1+SP+(3), where rule AN only applies to the axioms of S1 and to (3) (but not to the axioms of propositional identity SP). Finally, we observe that adding the scheme

\noindent (4) \centerline{$\square\varphi\rightarrow\square\square\varphi$}

\noindent to the non-normal modal system S3 such that rule AN also applies to the axioms (4) results in the normal modal logic S4. Recall that we only work with rule AN instead of the full Necessitation Rule: ``If $\vdash\psi$, then $\vdash\square\psi$." This stronger rule, however, is derivable in S1+SP+(3)+(4). We show this in a similar way as in \cite{lew}, Lemma 2.5, by induction on the length of the derivation of $\psi$ in system S1+SP+(3)+(4)=S3+(4). If $\psi$ is an axiom of the form (i)--(iii), (3) or (4), then we may apply AN. If $\psi$ is an axiom of the form (iv)--(vi), then we may apply Lemma \ref{420}. In any case, we conclude that $\square\psi$ is a theorem. If $\psi$ is the result of an application of AN to an axiom $\varphi$, then $\psi=\square\varphi$. By (4) and MP we get $\square\square\varphi$, i.e., $\square\psi$. Finally, let $\psi$ be the result of an application of MP to theorems $\varphi\rightarrow\psi$ and $\varphi$. By induction hypothesis, $\square(\varphi\rightarrow\psi)$ and $\square\varphi$ are theorems. Since modal law K is a theorem of S1+SP, we derive $\square\psi$. 

\begin{corollary}
S4=S3+(4).
\end{corollary}

It is well-known that S5=S4+(5), where the axiom scheme (5) is given by

\noindent (5) \centerline{$\neg\square\varphi\rightarrow\square\neg\square\varphi$.}\\

\noindent We have the following hierarchy of ``Lewis-style" modal systems:

\centerline{S1+SP $\subsetneq$ S1+SP+(3)=S3 $\subsetneq$ S3+(4)=S4 $\subsetneq$ S4+(5)=S5.}

\noindent Let us look at the semantic counterparts. The class of all models satisfying the additional condition

\noindent (3') \centerline{$f_\square (f_\rightarrow(m,m'))\le f_\square(f_\rightarrow(f_\square(m), f_\square(m')))$} 

\noindent generates a semantics for which S3 is sound and complete. In fact, condition (3') corresponds to the theorem that results from an application of AN to principle (3). 

\noindent We strengthen condition (iv) of Definition \ref{100} to

\noindent (4') \centerline{$f_\square(m)=f_\top\Leftrightarrow m=f_\top$,} 

\noindent which is the semantic counterpart of axiom scheme (4). That is, S4 is sound and complete with respect to the semantics determined by the constraints (3') and (4'). Finally, it is not hard to show that models which have property (4') and satisfy axiom scheme (5) also satisfy the condition 

\noindent (5') \centerline{$f_\square(m)\neq f_\top\Leftrightarrow f_\neg(f_\square(m))=f_\top$.} 

\noindent This can be shorter written as $f_\square(m)\neq f_\top\Leftrightarrow f_\square(m)=f_\bot$ (apply the operation of negation to both sides of the second equation). 

We would like to point out that models satisfying (3')+(4') also satisfy the conditions $f_\square(f_\top)=f_\top$ and $f_\square(f_\wedge(m,m'))=f_\wedge(f_\square(m),f_\square(m'))$, for all $m,m'$. The former equation follows readely from (4') and the latter one follows from the fact that $\square(\square(\varphi\wedge\psi)\leftrightarrow(\square\varphi\wedge\square\psi))$ is a theorem of S4. Thus, models satisfying (3')+(4') are specific \textit{modal algebras}.\\

Now it is clear how to modify our original completeness proof in order to obtain the following results.

\begin{corollary}
S3 is sound and complete w.r.t. the semantics generated by the class of all models that satisfy the condition (3'). S4 is sound and complete w.r.t. the semantics given by the class of all models that satisfy (3') and (4'). Finally, S5 is sound and complete w.r.t. the class of models that satisfy the constraints (3'), (4') and (5').
\end{corollary}

Of course, we get \textit{strong} completeness results such as formulated in Theorem \ref{360}. We conclude that our approach provides an uniform semantical framework for the ``Lewis-like" modal logic S1+SP and the Lewis systems S3--S5. By Theorem \ref{400}, the systems S1 and S2 cannot be captured. If we enlarge the language by a connective $\equiv$, then S1+SP augmented with the axiom scheme $(\varphi\equiv\psi)\leftrightarrow\square(\varphi\leftrightarrow\psi)$ can be seen as a minimal amalgam of Suszko's basic non-Fregean logic SCI \cite{blosus} and Lewis modal logic S1.

\end{document}